# A Lax Pair for the Dynamics of DNA Modeled as a Shearable and Extensible Elastic Rod: III. Discretization of the Arc Length and Time

Yaoming Shi(a), W.M. McClain(b), and John E. Hearst(a)

(a)Department of Chemistry, University of California, Berkeley, CA 94720-1460
(b)Department of Chemistry, Wayne State University, Detroit, MI 48202

Contact: jehearst@cchem.berkeley.edu

July 24, 2003

## ABSTRACT

We find a Lax pair for the geometrically exact discrete Hamiltonian equations for the discrete elastic rod. This is paper III in a series.

# I. INTRODUCTION

In paper I[1], we derived a Lax Pair for the geometrically exact Hamiltonian equations for the elastic rod. In that paper, two independent parameters, t and s, were considered to be continuous. In paper II[2], we provided a parallel treatment for the case in which t is continuous and s is discrete. In this paper, we derive a Lax Pair for the t-discrete and s-discrete case.

# II. THE CONFIGURATION SPACE OF THE ELASTIC ROD IN 3D

We treat duplex DNA as a bendable, twistable, extensible, and shearable thin elastic rod. Here and elsewhere in this paper the terms "elastic rod" and "DNA" have the same meaning; so do "the centerline of the rod" and "the axis of DNA".

Let $l$ and $k$ be two integers. In order to simplify our notations, we will express any function $f(s,t)$ as $f(k\Delta s, l\Delta t) = f(k,l)$. At a given time $t = l\,\Delta t$ and at each point $s = k\,\Delta s$ on the centerline $\mathbf{r}(k,l)$ of the rod, a localized Cartesian coordinate frame (or director frame), $\{\hat{\mathbf{d}}_1(k,l), \hat{\mathbf{d}}_2(k,l), \hat{\mathbf{d}}_3(k,l)\}$, is affixed with the unit vectors $\hat{\mathbf{d}}_1(k,l)$ and $\hat{\mathbf{d}}_2(k,l)$ in the direction of the principal axes of inertia tensor of the rod cross section. The third unit vector $\hat{\mathbf{d}}_3(k,l)$ is in the normal direction of the cross section. Because the shear is included, the unit vector $\hat{\mathbf{d}}_3(k,l)$ does not necessarily coincide with the tangent vector $\Gamma(k,l)$ of the centerline of the elastic rod $\mathbf{r}(k,l)$.

Unit vectors $\hat{\mathbf{d}}_a(k,l)$ in the director frame (or body-fixed frame) are related to the unit vectors $\hat{\mathbf{a}}_a$ in the lab frame via an Euler rotation matrix $\Lambda$ according to
$$\hat{\mathbf{d}}_a(k,l) = \Lambda(\varphi(k,l), \theta(k,l), \psi(k,l)) \cdot \hat{\mathbf{a}}_a = \Lambda(k,l) \cdot \hat{\mathbf{a}}_a, \qquad (a = 1,2,3). \tag{2.1}$$
Since $\mathbf{r} \in R^3$ and $\Lambda \in SO(3)$, the configuration space of the elastic rod is $E^3 = R^3 \times SO(3)$.

The orientation of the local frame at $s + \Delta s = (k+1)\Delta s$ is obtained by a tiny rotation of the coordinate frame at $s = k\Delta s$. The velocity of the rotation is the matrix
$$\overline{\Omega}(k,l) = [\Lambda(k,l)]^T \cdot [\Lambda(k+1,l) - \Lambda(k,l)]/\Delta s. \tag{2.2}$$
It can be shown that
$$\overline{\Omega}_{bc}(k,l) = \varepsilon_{abc}\Omega_a(k,l) + O(\Delta s) \tag{2.3}$$

where $\Omega_a(k,l)$ are the local components of the Darboux vector in the rod frame, namely
$$\mathbf{\Omega}(k,l) = \Omega_a(k,l)\hat{\mathbf{d}}_a(k,l). \tag{2.4}$$
Here and after, double occurrence of an index in the subscript means summation over its range. Vectors $\hat{\mathbf{d}}_a(k,l)$ change with $k$ by moving perpendicular to themselves and to the rotation axis $\mathbf{\Omega}(k,l)$ according to
$$\hat{\mathbf{d}}_a(k+1,l) = \hat{\mathbf{d}}_a(k,l) + \Delta s\, \mathbf{\Omega}(k,l) \times \hat{\mathbf{d}}_a(k,l) + O(\Delta s^2). \tag{2.5}$$
The relative position of the origin of the localized rod frame at $s+\Delta s = (k+1)\Delta s$ is obtained by a tiny translation of the origin of the localized frame at $s = k\Delta s$, i.e.,
$$\mathbf{r}(k+1,l) = \mathbf{r}(k,l) + \Delta s\, \mathbf{\Gamma}(k,l) + O(\Delta s^2). \tag{2.6}$$
The velocity of the translation is the tangent vector $\mathbf{\Gamma}(k,l)$ with local components
$$\mathbf{\Gamma}(k,l) = \Gamma_a(k,l)\hat{\mathbf{d}}_a(k,l). \tag{2.7}$$

The parameter $s = k\Delta s$, usually chosen as the arclength parameter for the undeformed (or relaxed) elastic rod, is no longer the current arclength parameter for the deformed rod, $\tilde{s}(s,t)$, since there are deformations of shear and extension. The current arclength of the deformed rod, $\tilde{s}(s,t)$, is then given by
$$\tilde{s}(k,l) = \Delta s \sum_{n=1}^{k} |\mathbf{\Gamma}(n,l)|.$$

The orientation of the local frame at time $t+\Delta t = (l+1)\Delta t$ is obtained by an infinitesimal rotation of the coordinate frame at time $t = l\Delta t$. The velocity of the rotation is the matrix
$$\overline{\omega}(k,l) = (\Delta t)^{-1} [\Lambda(k,l)]^T \cdot [\Lambda(k,l+1) - \Lambda(k,l)]. \tag{2.8}$$
It can be shown that
$$\overline{\omega}_{ab}(k,l) = \varepsilon_{abc}\omega_c(k,l) + O(\Delta t) \tag{2.9}$$
where $\omega_c(k,l)$ are the local frame components of the angular velocity vector $\mathbf{\omega}(k,l)$, namely,
$$\mathbf{\omega}(k,l) = \omega_c(k,l)\hat{\mathbf{d}}_c(k,l), \tag{2.10}$$
obeying
$$\hat{\mathbf{d}}_a(k,l+1) = \hat{\mathbf{d}}_a(k,l) + \Delta t\, \mathbf{\omega}(k,l) \times \hat{\mathbf{d}}_a(k,l) + O(\Delta t^2). \tag{2.11}$$

The relative position of the origin of the localized rod frame at $t+\Delta t = (l+1)\Delta t$ is obtained by a tiny translation of the origin of the localized frame at $t = l\Delta t$. The velocity of the translation is the linear velocity vector $\mathbf{\gamma}(k,l)$:
$$\mathbf{r}(k,l+1) = \mathbf{r}(k,l) + \Delta t\, \mathbf{\gamma}(k,l) + O(\Delta t^2) \tag{2.12}$$
$$\mathbf{\gamma}(k,l) = \gamma_a(k,l)\hat{\mathbf{d}}_a(k,l). \tag{2.13}$$

Three dependent variables are used for describing strains (or deformations) of bending$_1$ [$\Omega_1(k,l)$], bending$_2$ [$\Omega_2(k,l)$], and twisting [$\Omega_3(k,l)$]. Another three are used for describing the strains (or deformations) of shear$_1$ [$\Gamma_1(k,l)$], shear$_2$ [$\Gamma_2(k,l)$], and extension [$\Gamma_3(k,l)$].

For DNA, these strains are related the relative motion between adjacent base pairs. For example, if we pick $\Delta s = 0.328 nm$, to be the distance between DNA base pairs, then quantities $\Omega_1(k,l)\Delta s = Roll$, $\Omega_2(k,l)\Delta s = Tilt$, $\Omega_3(k,l)\Delta s = Twist$, $\Gamma_1(k,l)\Delta s = Slide$, $\Gamma_2(k,l)\Delta s = Shift$, and $\Gamma_3(k,l)\Delta s = Rise$. Figure 1[3] in Appendix A shows 6 drawings of these six kinds of motions between adjacent base pairs.

Still another three are used for describing the linear velocity$_1$ [$\gamma_1(k,l)$], linear velocity$_2$ [$\gamma_2(k,l)$], and linear velocity$_3$ [$\gamma_3(k,l)$] for the translation of the centroid of the elastic rod cross section at position $s = k\Delta s$ and time $t = l\Delta t$. The last three are used for describing the angular velocity$_1$ [$\omega_1(k,l)$], angular velocity$_2$ [$\omega_2(k,l)$], and angular velocity$_3$ [$\omega_3(k,l)$] for the rotation of the elastic rod cross section at position $s = k\Delta s$ and time $t = l\Delta t$.

The stresses (internal torques and forces) corresponding to the strains are denoted by dependent variables $M_1(k,l)$, $M_2(k,l)$, $M_3(k,l)$, $P_1(k,l)$, $P_2(k,l)$, and $P_3(k,l)$. Since each cross section has its mass and moment of inertia tensor, an angular momentum $\mathbf{m}(k,l)$ of the cross section and a linear momentum $\mathbf{p}(k,l)$ of the center of the cross section can be naturally introduced. The local components of the angular momentum are $m_1(k,l)$, $m_2(k,l)$, $m_3(k,l)$, and the local components of the linear momentum are $p_1(k,l)$, $p_2(k,l)$, $p_3(k,l)$.

From this point on, the letters
$\Omega(k,l)$, $\Gamma(k,l)$, $M(k,l)$, $P(k,l)$, $\omega(k,l)$, $\gamma(k,l)$, $m(k,l)$, $p(k,l)$, without subscripts will be understood as vectors, whether bold or not. The plane face letter $\Omega$ means $\Omega = (\Omega_1, \Omega_2, \Omega_3)^T$ and the bold face letter $\boldsymbol{\Omega}$ means $\boldsymbol{\Omega} = \Omega_1 \hat{\mathbf{d}}_1 + \Omega_2 \hat{\mathbf{d}}_2 + \Omega_3 \hat{\mathbf{d}}_3$.

## III. LAX PAIR FOR THE DISCRETE SMK EQUATIONS

Let $J_a$ and $K_a$ ($a = 1, 2, 3$) be the generator matrices of Lie group $SO(4)$, given by

$$J_1 = \begin{pmatrix} 0 & 0 & 0 & 0 \\ 0 & 0 & -1 & 0 \\ 0 & 1 & 0 & 0 \\ 0 & 0 & 0 & 0 \end{pmatrix}, \quad J_2 = \begin{pmatrix} 0 & 0 & 1 & 0 \\ 0 & 0 & 0 & 0 \\ -1 & 0 & 0 & 0 \\ 0 & 0 & 0 & 0 \end{pmatrix}, \quad J_3 = \begin{pmatrix} 0 & -1 & 0 & 0 \\ 1 & 0 & 0 & 0 \\ 0 & 0 & 0 & 0 \\ 0 & 0 & 0 & 0 \end{pmatrix} \quad (3.1a)$$

$$K_1 = \begin{pmatrix} 0 & 0 & 0 & -1 \\ 0 & 0 & 0 & 0 \\ 0 & 0 & 0 & 0 \\ 1 & 0 & 0 & 0 \end{pmatrix}, \quad K_2 = \begin{pmatrix} 0 & 0 & 0 & 0 \\ 0 & 0 & 0 & -1 \\ 0 & 0 & 0 & 0 \\ 0 & 1 & 0 & 0 \end{pmatrix}, \quad K_3 = \begin{pmatrix} 0 & 0 & 0 & 0 \\ 0 & 0 & 0 & 0 \\ 0 & 0 & 0 & -1 \\ 0 & 0 & 1 & 0 \end{pmatrix}. \quad (3.1b)$$

These generators satisfy the relations
$$[J_a, J_b] = \varepsilon_{abc} J_c, \quad [J_a, K_b] = \varepsilon_{abc} K_c, \quad [K_a, K_b] = \varepsilon_{abc} J_c \quad (3.1c)$$
where $\varepsilon_{abc}$ is the Levi-Civita symbol.

In order to simplify the notation further, we will also express symbol $f_a(k,l,\lambda)$ as $f_a(k,l,\lambda) = f_{a,0}^0(k,l,\lambda) = f_a$ and symbol $f(k+1,l,\lambda)$ as $f_a(k+1,l,\lambda) = f_a^1(k,l,\lambda) = f_a^1$ and symbol $f_a(k,l+1,\lambda)$ as $f_a(k,l+1,\lambda) = f_{a,1}(k,l,\lambda) = f_{a,1}$.

Now consider the linear system
$$\Phi^1 = \tilde{U} \, \Phi \quad (3.2a)$$
$$\Phi_1 = \tilde{V} \, \Phi \quad (3.2b)$$
where $V$ and $V$ are defined as
$$\tilde{U} = \mathbf{1}_4 + \Delta s \, U \quad (3.3a)$$
$$\tilde{V} = \mathbf{1}_4 + \Delta t \, V \quad (3.3b)$$
$$U = A + \sqrt{-1} \, B \quad (3.3c)$$
$$V = C + \sqrt{-1} \, D \quad (3.3d)$$
and where $\mathbf{1}_4$ is the 4-by-4 identity matrix and $A, B, C, D$ are given by
$$A = -\left( \Omega_a J_a + \lambda^2 \Gamma_a K_a \right) \quad (3.4)$$
$$B = -\lambda \left( p_a J_a + \lambda^2 m_a K_a \right) \quad (3.5)$$
$$C = -\left( \omega_a J_a + \lambda^2 \gamma_a K_a \right) \quad (3.6)$$
$$D = -\lambda \left( P_a J_a + \lambda^2 M_a K_a \right). \quad (3.7)$$

In system (3.4-3.7), symbols $\Omega_a$, $\Gamma_a$, $M_a$, $P_a$, $\omega_a$, $\gamma_a$, $m_a$, $p_a$, are real functions of $k$, $l$ and $\lambda$, the real spectral parameter.

The integrability condition for the linear system (3.2) is then given by:
$$\tilde{U}_1 \, \tilde{V} = \tilde{V}^1 \, \tilde{U}. \quad (3.8)$$

Substituting (3.3) into (3.8) and separating the real part from the imaginary part, we obtain:

$$(\Delta t)^{-1}\left(\Gamma_{c,1}-\Gamma_c\right)-(\Delta s)^{-1}\left(\gamma_c^1-\gamma_c\right)$$
$$-\varepsilon_{abc}\left(\Gamma_a\,\omega_b^1+\Omega_{a,1}\,\gamma_b\right)-(\lambda)^2\,\varepsilon_{abc}\left(M_{a,1}\,p_{b,1}-m_a\,P_b^1\right)=0 \tag{3.9a}$$

$$(\Delta t)^{-1}\left(\Omega_{a,1}-\Omega_a\right)-(\Delta s)^{-1}\left(\omega_a^1-\omega_a\right)$$
$$-\left(\Omega_{a+1}\,\omega_{a+2}^1-\omega_{a+1}\,\Omega_{a+2,1}\right)-(\lambda)^2\left(P_{a+1}\,p_{a+2,1}-p_{a+1}\,P_{a+2}^1\right) \tag{3.9b}$$
$$+(\lambda)^4\left(\Gamma_{a+1,1}\,\gamma_{a+2}-\gamma_{a+1}^1\,\Gamma_{a+2}\right)+(\lambda)^6\left(M_{a+1}^1\,m_{a+2}-m_{a+1,1}\,M_{a+2}\right)=0$$

$$(\Delta t)^{-1}\left(p_{a,1}-p_a\right)-(\Delta s)^{-1}\left(P_a^1-P_a\right)$$
$$-\left(p_{a+1}\,\omega_{a+2}^1-\omega_{a+1}\,p_{a+2,1}\right)+\left(P_{a+1}\,\Omega_{a+2,1}-\Omega_{a+1}\,P_{a+2}^1\right) \tag{3.9c}$$
$$+(\lambda)^4\left(\Gamma_{a+1}\,M_{a+2}^1-M_{a+1}\,\Gamma_{a+2,1}\right)+(\lambda)^4\left(\gamma_{a+1}^1\,m_{a+2}-m_{a+1,1}\,\gamma_{a+2}\right)=0$$

$$(\Delta t)^{-1}\left(m_{c,1}-m_c\right)-(\Delta s)^{-1}\left(M_c^1-M_c\right)$$
$$-\varepsilon_{abc}\left(p_{a,1}\,\gamma_b+\Gamma_a\,P_b^1+\Omega_{a,1}\,M_b+m_a\,\omega_b^1\right)=0 \tag{3.9d}$$

In (3.9) above, index $a$ runs 1,2,3. Index $a+1$ and index $a+2$ are assumed to take the values of modula 3.

In Appendix B we provide a different derivation of Eqs.(3.9a-d) using a different Lax pair.

When $\Delta s=\Delta t\to 0$, $k\to\infty$, $l\to\infty$, but $k\Delta s=(k+1)\Delta s=s$ and $l\Delta t=(l+1)\Delta t=t$ remain finite, system (3.9) reduces to its s-t-continuous counterpart [(3.5) in Paper I[4]].

When $\Delta t\to 0$, $l\to\infty$, but $l\Delta t=(l+1)\Delta t=t$ remain finite, system (3.9) reduces to its t-continuous and s-discrete counterpart [Eq.(3.9) in paper II[5]].

Since system (3.9a-d) contains 12 scalar equations for 24 real dependent variables, $\Omega_a$, $\Gamma_a$, $\omega_a$, $\gamma_a$, $M_a$, $P_a$, $m_a$, $p_a$ $(a=1,2,3)$, we have the freedom to pick twelve real "constitutive" relations. In elastic work one uses the twelve scalar equations implied by the four vector equations

$$P_a=\frac{\partial}{\partial\Gamma_a}H(\Omega,\Gamma) \tag{3.10a}$$

$$M_a=\frac{\partial}{\partial\Omega_a}H(\Omega,\Gamma) \tag{3.10b}$$

$$p_a = \frac{\partial}{\partial \gamma_a} h(\omega,\gamma,t) \tag{3.10c}$$

$$m_a = \frac{\partial}{\partial \omega_a} h(\omega,\gamma,t) \tag{3.10d}$$

where $(a = 1,2,3)$, $H(\Omega,\Gamma)$ is the elastic energy function, and $h(\omega,\gamma)$ is the kinetic energy function. But in other problems one might use other relations, which we write symbolically as

$$H_A(\Omega,\Gamma,M,P,\omega,\gamma,m,p,\lambda) = 0, \qquad (A=1,2,\ldots, 12). \tag{3.11}$$

System (3.9) and (3.11) can describe a very large class of Partial-Difference-Equations [24 dependent variables in (1+1) dimension]. Some members of this class will be discussed in detail in Section IV, below.

What is new about this Lax pair and the resulting system of 12 Partial-Difference-Equations? The pure kinematic approach[6] for recasting a the nonlinear Partial-Difference-Equations of a general curve into Lax representation focuses only on what we call the strain-velocity compatibility (integrability) relations for equations like

$$\Phi' = Q(\Omega, \Gamma, \lambda)\Phi \qquad \text{and} \qquad \Phi_1 = R(\omega, \gamma, \lambda)\Phi$$

where $\Omega$ and $\Gamma$ are strains and $\omega$ and $\gamma$ are velocities. But we treat strain-velocity and stress-momentum on an equal footing, so our equations are

$$\Phi' = U(\Omega, \Gamma, p, m, \lambda)\Phi \qquad \text{and} \qquad \Phi_1 = V(\omega, \gamma, P, M, \lambda)\Phi$$

where the new variables $P$ and $M$ are stresses (force and torque, respectively) and $p$ and $m$ are momenta (linear and angular, respectively). The Lax equations for the two cases look identical

$$Q_1 R = R' Q \qquad \textit{versus} \qquad U_1 V = V' U,$$

and quantities $Q$, $R$, $U$, and $V$ are all 4-by-4 arrays, but $Q$ and $R$ are real, whereas our $U$ and $V$ are complex and contain twice as many dependent variables.

We emphasize that the stress-momenta appear as naturally as the strain-velocities, and the resulting nonlinear Partial-Difference-Equations are dynamic elastic equations rather than just kinematic equations.

Expanding the dependent variables $X_a = X_a(\lambda)$ and the constitutive relations $H_A(\Omega,\Gamma,M,P,\omega,\gamma,m,p,\lambda)$ in Taylor series in $\lambda$,

$$X_a(\lambda) = \sum_{n=0}^{\infty} \lambda^n X_a^{(n)} \text{ and } H_A(\lambda) = \sum_{n=0}^{\infty} \lambda^n H_A^{(n)}, \tag{3.12}$$

and taking the limit $\lambda \to 0$, we find that the leading terms in the expansions (3.9) are the s-and-t-discrete SMK equations:

$$\begin{aligned}(\Delta t)^{-1}\left(\Gamma_{c,1} - \Gamma_c\right) - (\Delta s)^{-1}\left(\gamma_c^1 - \gamma_c\right) \\ -\varepsilon_{abc}\left(\Gamma_a \omega_b^1 + \Omega_{a,1} \gamma_b\right) = 0\end{aligned} \tag{3.13a}$$

$$(\Delta t)^{-1}\left(\Omega_{a,1} - \Omega_a\right) - (\Delta s)^{-1}\left(\omega_a^1 - \omega_a\right) \\ - \left(\Omega_{a+1}\,\omega_{a+2}^1 - \omega_{a+1}\,\Omega_{a+2,1}\right) = 0 \qquad (3.13b)$$

$$(\Delta t)^{-1}\left(p_{a,1} - p_a\right) - (\Delta s)^{-1}\left(P_a^1 - P_a\right) \\ - \left(p_{a+1}\,\omega_{a+2}^1 - \omega_{a+1}\,p_{a+2,1}\right) + \left(P_{a+1}\,\Omega_{a+2,1} - \Omega_{a+1}\,P_{a+2}^1\right) = 0 \qquad (3.13c)$$

$$(\Delta t)^{-1}\left(m_{c,1} - m_c\right) - (\Delta s)^{-1}\left(M_c^1 - M_c\right) \\ - \varepsilon_{abc}\left(p_{a,1}\,\gamma_b + \Gamma_a\,P_b^1 + \Omega_{a,1}\,M_b + m_a\,\omega_b^1\right) = 0 \qquad (3.13d)$$

In (3.13) above, index $a$ runs 1,2,3. Index $a+1$ and index $a+2$ are assumed to take the values of modula 3.

When $\Delta s = \Delta t \to 0$, $k \to \infty$, $l \to \infty$, but $k\Delta s = (k+1)\Delta s = s$ and $l\Delta t = (l+1)\Delta t = t$ remain finite, System (3.13) reduces to its s-continuous and t-continuous SMK equations [(2.1) in paper I[7]].

When $\Delta t \to 0$, $l \to \infty$, but $l\Delta t = (l+1)\Delta t = t$ remain finite, System (3.13) reduces to its t-continuous and s-discrete SMK equations [(3.13)) in paper II[8]].

# IV. SPECIALIZATION OF THE SMK EQUATIONS

## Case 1. The Static Elastic Rod

Setting velocities and momenta $\omega_a = \gamma_a = m_a = p_a = 0$ in (3.13) and assuming that everything else is independent variable $l$ or $l+1$, we find that the s-and-t-discrete SMK equations (3.13) reduce to the following difference equations:

$$(\Delta s)^{-1}\left(P_a^1 - P_a\right) - \left(P_{a+1}\,\Omega_{a+2} - \Omega_{a+1}\,P_{a+2}^1\right) = 0 \qquad (4.1a)$$

$$(\Delta s)^{-1}\left(M_c^1 - M_c\right) + \varepsilon_{abc}\left(\Gamma_a\,P_b^1 + \Omega_a\,M_b\right) = 0 \qquad (4.1b)$$

$$P_a = \frac{\partial}{\partial \Gamma_a} H(\Omega,\Gamma) \qquad (4.1c)$$

$$M_a = \frac{\partial}{\partial \Omega_a} H(\Omega,\Gamma). \qquad (4.1d)$$

System (4.1) describes the equilibrium configurations of a discrete elastic rod with elastic energy function $H(\Omega,\Gamma)$.

### Subcase 1.1  Static Elastic Rod with Linear Constitutive Relations

Let $H(\Omega, \Gamma)$ of (4.1c) and (4.1d) be given by

$$H(\Omega,\Gamma) = \tfrac{1}{2} A_{ab} \left(\Omega_a - \Omega_a^{(\text{intrinsic})}\right)\left(\Omega_b - \Omega_b^{(\text{intrinsic})}\right)$$
$$+ \tfrac{1}{2} C_{ab}\left(\Gamma_a - \Gamma_a^{(\text{intrinsic})}\right)\left(\Gamma_b - \Gamma_b^{(\text{intrinsic})}\right) +$$
$$+ \tfrac{1}{2} B_{ab}\left[\left(\Omega_a - \Omega_a^{(\text{intrinsic})}\right)\left(\Gamma_a - \Gamma_b^{(\text{intrinsic})}\right)\right.$$
$$\left. + \left(\Gamma_a - \Gamma_a^{(\text{intrinsic})}\right)\left(\Omega_b - \Omega_b^{(\text{intrinsic})}\right)\right]$$
(4.2)

where $A_{ab}$ is the bending/twisting modulus, $C_{ab}$ is the shear/extension modulus, and $B_{ab}$ is the coupling modulus between bending/twisting and shear/extension. The quantities $\Omega_a^{(\text{intrinsic})}$ are the intrinsic bending and twisting of the unstressed rod, and the quantities $\Gamma_a^{(\text{intrinsic})}$ are the intrinsic shear and extension.

For DNA, the values of these intrinsic strains and modulus can be chosen [9] from Table I and Table II in Appendix A.

Then system (4.1) reduces to
$$(\Delta s)^{-1}\left(P_a^1 - P_a\right) - \left(P_{a+1}\,\Omega_{a+2} - \Omega_{a+1}\,P_{a+2}^1\right) = 0 \tag{4.3a}$$
$$(\Delta s)^{-1}\left(M_c^1 - M_c\right) + \varepsilon_{abc}\left(\Gamma_a\,P_b^1 + \Omega_a\,M_b\right) = 0 \tag{4.3b}$$
$$P_a = C_{ab}\left(\Gamma_b - \Gamma_b^{(\text{intrinsic})}\right) + B_{ab}\left(\Omega_b - \Omega_b^{(\text{intrinsic})}\right). \tag{4.3c}$$
$$M_a = A_{ab}\left(\Omega_b - \Omega_b^{(\text{intrinsic})}\right) + B_{ab}\left(\Gamma_b - \Gamma_b^{(\text{intrinsic})}\right) \tag{4.3d}$$

The s-continuous limit of the system (4.3a,b) in this paper is identical to system (4.3a,b) in paper I[10]. We have shown in paper I that, when
(1) $B_{ab} = 0$ and
(2) $A_{ab}$ is a constant diagonal matrix with $A_{11} = A_{22}$, and
(3) $C_{ab}$ is a constant diagonal matrix with $C_{11} = C_{22}$, and
(4) $\Omega_a^{(\text{intrinsic})} = 0$, $\Gamma_1^{(\text{intrinsic})} = \Gamma_2^{(\text{intrinsic})} = \Gamma_3^{(\text{intrinsic})} - 1 = 0$,

the s-continuous limit of the system (4.3a,b) may be solved exactly in terms of elliptic functions[11]. Open question: when the above conditions pertain, is this difference system (4.3a,b) also integrable?

### Subcase 1.2   Static Kirchhoff Elastic Rod, and the Heavy Top

Assuming that the elastic rod does not have shear and extension deformations (i.e., $\Gamma_a = \Gamma_a^{(\text{intrinsic})}$, $\Gamma_1^{(\text{intrinsic})} = \Gamma_2^{(\text{intrinsic})} = \Gamma_3^{(\text{intrinsic})} - 1 = 0$), then $H(\Omega, s)$ becomes
$$H(\Omega) = \tfrac{1}{2} A_{ab}\left(\Omega_a - \Omega_a^{(\text{intrinsic})}\right)\left(\Omega_b - \Omega_b^{(\text{intrinsic})}\right) \tag{4.4a}$$
and system (4.3) reduces to

$$(\Delta s)^{-1}(P_a^1 - P_a) - (P_{a+1}\Omega_{a+2} - \Omega_{a+1}P_{a+2}^1) = 0 \tag{4.5a}$$

$$(\Delta s)^{-1}(M_c^1 - M_c) + \varepsilon_{abc}(\Gamma_a P_b^1 + \Omega_a M_b) = 0 \tag{4.5b}$$

$$M_a = A_{ab}(\Omega_b - \Omega_b^{(\text{intrinsic})}). \tag{4.5c}$$

If $s = k\Delta s$ is arc length (as assumed earlier), then system (4.5) describes the equilibrium configuration of the unshearable, inextensible discrete Kirchhoff elastic rod. But if $s = k\Delta s$ is understood as time, this system describes the time-discrete dynamics of the heavy top.

There are two known integrable cases for the time-continuous version of the heavy top system (4.5) with $\Omega^{(\text{intrinsic})} = (0,0,0)^T$, $\Gamma = (0,0,1)^T$:

(a) Lagrange Top[12]: $A_{ab}$ is a constant diagonal matrix with $A_{11} = A_{22}$.

(b) Kowalewski Top[13]: the same, but with $A_{33} = A_{11} = 2A_{22}$.

Open question: what about the time-discrete dynamics of the Lagrange top and the Kowalewski top?

## Case 2, Rigid Body Motion In Ideal Fluid

Setting $\Omega_{c,1} = \Omega_c = 0$, $\Gamma_{c,1} = \Gamma_c = 0$, $M_{c,1} = M_c = 0$, $P_{c,1} = P_c = 0$, in (3.13) and assuming that everything else is a function of time $t = l\Delta t$ only, then $\gamma_c^1 = \gamma_c$, $\omega_c^1 = \omega_c$, and the SMK equations (3.13) reduce to

$$(\Delta t)^{-1}(p_{a,1} - p_a) - (p_{a+1}\omega_{a+2}^1 - \omega_{a+1}p_{a+2,1}) = 0. \tag{5.1a}$$

$$(\Delta t)^{-1}(m_{c,1} - m_c) - \varepsilon_{abc}(p_{a,1}\gamma_b + m_a \omega_b^1) = 0 \tag{5.1b}$$

$$p_a = \frac{\partial}{\partial \gamma_a} h(\omega,\gamma) \tag{5.1c}$$

$$m_a = \frac{\partial}{\partial \omega_a} h(\omega,\gamma). \tag{5.1d}$$

In (5.1) above, index $a$ runs 1,2,3. Index $a+1$ and index $a+2$ are assumed to take the values of modula 3.

When $\Delta t \to 0$, $l \to \infty$, but $l\Delta t = (l+1)\Delta t = t$ remain finite, System (5.1) reduces to its t-continuous counterpart [(5.1) in paper I[14]]. In paper I, we also discussed three known integrable cases; namely, the Clebsch case, the Steklov case, and the Chaplygin case.

## Case 3, Kirchhoff Elastic Rod Motion

In this case, there is no shear or extension, (i.e., $\Gamma_a = \Gamma_a^{(\text{intrinsic})}, \Gamma_1^{(\text{intrinsic})} = \Gamma_2^{(\text{intrinsic})} = \Gamma_3^{(\text{intrinsic})} - 1 = 0$), and the SMK equations in (3.13) reduce to:

$$(\Delta s)^{-1}(\gamma_c^1 - \gamma_c) + \varepsilon_{abc}(\Gamma_a \omega_b^1 + \Omega_{a,1} \gamma_b) = 0 \tag{6.1a}$$

$$(\Delta t)^{-1}(\Omega_{a,1} - \Omega_a) - (\Delta s)^{-1}(\omega_a^1 - \omega_a) \\ - (\Omega_{a+1} \omega_{a+2}^1 - \omega_{a+1} \Omega_{a+2,1}) = 0 \tag{6.1b}$$

$$(\Delta t)^{-1}(p_{a,1} - p_a) - (\Delta s)^{-1}(P_a^1 - P_a) \\ - (p_{a+1} \omega_{a+2}^1 - \omega_{a+1} p_{a+2,1}) + (P_{a+1} \Omega_{a+2,1} - \Omega_{a+1} P_{a+2}^1) = 0 \tag{6.1c}$$

$$(\Delta t)^{-1}(m_{c,1} - m_c) - (\Delta s)^{-1}(M_c^1 - M_c) \\ - \varepsilon_{abc}(p_{a,1} \gamma_b + \Gamma_a P_b^1 + \Omega_{a,1} M_b + m_a \omega_b^1) = 0 \tag{6.1d}$$

In (6.1) above, index $a$ runs 1,2,3. Index $a+1$ and index $a+2$ are assumed to take the values of modula 3.

System (6.1) describes the dynamics of the discrete Kirchhoff elastic if we pick

$$M_a = A_{ab}(\Omega_b - \Omega_b^{(\text{intrinsic})}) \tag{6.1e}$$
$$m_a = I_{ab} \omega_b \tag{6.1f}$$
$$p_a = \rho \gamma_a \tag{6.1g}$$

where $I_{ab}$ is the moment of inertia tensor for the cross section of the elastic rod and $\rho$ is the linear density of mass.

## Case 4, Elastic Rod Moving in a Plane

Setting $\omega_1 = \omega_2 = m_1 = m_2 = \gamma_3 = p_3 = 0$ and $\Omega_1 = \Omega_2 = M_1 = M_2 = \Gamma_3 = P_3 = 0$ in (3.13), then we find that the SMK equations reduce to:

$$(\Delta t)^{-1}(\Gamma_{1,1} - \Gamma_1) - (\Delta s)^{-1}(\gamma_1^1 - \gamma_1) \\ + \Gamma_2 \omega_3^1 - \Omega_{3,1} \gamma_{2,} = 0 \tag{7.1a}$$

$$(\Delta t)^{-1} (\Gamma_{2,1} - \Gamma_2) - (\Delta s)^{-1} (\gamma_2^1 - \gamma_2)$$
$$-\Gamma_1 \omega_3^1 + \Omega_{3,1} \gamma_1 = 0 \tag{7.1b}$$

$$(\Delta t)^{-1} (\Omega_{3,1} - \Omega_3) - (\Delta s)^{-1} (\omega_3^1 - \omega_3) = 0 \tag{7.1c}$$

$$(\Delta t)^{-1} (m_{3,1} - m_3) - (\Delta s)^{-1} (M_3^1 - M_3)$$
$$+ p_{1,1} \gamma_2 - p_{2,1} \gamma_1 + \Gamma_1 P_2^1 - \Gamma_2 P_1^1 = 0 \tag{7.1d}$$

$$(\Delta t)^{-1} (p_{1,1} - p_1) - (\Delta s)^{-1} (P_1^1 - P_1)$$
$$+ p_2 \omega_3^1 - P_2 \Omega_{3,1} = 0 \tag{7.1e}$$

$$(\Delta t)^{-1} (p_{2,1} - p_2) - (\Delta s)^{-1} (P_2^1 - P_2)$$
$$- p_1 \omega_3^1 + P_1 \Omega_{3,1} = 0 \tag{7.1f}$$

System (7.1a – 7.1f) contains six scalar equations for twelve dependent variables. The six constitutive relations can be expressed as

$$P_1 = \frac{\partial}{\partial \Gamma_1} H(\Omega_3, \Gamma_1, \Gamma_2) \tag{7.2a}$$

$$P_2 = \frac{\partial}{\partial \Gamma_2} H(\Omega_3, \Gamma_1, \Gamma_2) \tag{7.2b}$$

$$M_3 = \frac{\partial}{\partial \Omega_3} H(\Omega_3, \Gamma_1, \Gamma_2) \tag{7.2c}$$

$$p_1 = \frac{\partial}{\partial \gamma_1} h(\omega_3, \gamma_1, \gamma_2) \tag{7.2d}$$

$$p_2 = \frac{\partial}{\partial \gamma_2} h(\omega_3, \gamma_1, \gamma_2) \tag{7.2e}$$

$$m_3 = \frac{\partial}{\partial \omega_3} h(\omega_3, \gamma_1, \gamma_2). \tag{7.2f}$$

In system (7.1), the dependent variable $\Omega_3$ is used for describing bending, $\Gamma_1$ for shear, $\Gamma_2$ for extension, $\gamma_1$ for linear velocity$_1$, $\gamma_2$ for linear velocity$_2$, and $\omega_3$ for angular velocity. In addition, $h(\Omega_3, \Gamma_1, \Gamma_2)$ stands for the elastic energy and $h(\omega_3, \gamma_1, \gamma_2)$ for the kinetic energy.

# Case 5, Moving Space Curve (in 3D)

The first two kinematic vector equations (3.13a, 3.13b) in the SMK equations (3.13) can be derived from the following discrete Serret-Frenet equations for an elastic rod moving on the 3-dimensional surface of a 4-dimensional sphere with radius $\lambda^{-2}$, namely,

$$\begin{pmatrix} \hat{\mathbf{d}}_1(k+1,l) \\ \hat{\mathbf{d}}_2(k+1,l) \\ \hat{\mathbf{d}}_3(k+1,l) \\ \hat{\mathbf{r}}(k+1,l) \end{pmatrix} = (\Delta s) \begin{pmatrix} (\Delta s)^{-1} & \Omega_3 & -\Omega_2 & -\lambda^2\, \Gamma_1 \\ -\Omega_3 & (\Delta s)^{-1} & -\Omega_1 & -\lambda^2\, \Gamma_2 \\ -\Omega_2 & \Omega_1 & (\Delta s)^{-1} & -\lambda^2\, \Gamma_3 \\ \lambda^2\, \Gamma_1 & \lambda^2\, \Gamma_2 & \lambda^2\, \Gamma_3 & (\Delta s)^{-1} \end{pmatrix} \begin{pmatrix} \hat{\mathbf{d}}_1(k,l) \\ \hat{\mathbf{d}}_2(k,l) \\ \hat{\mathbf{d}}_3(k,l) \\ \hat{\mathbf{r}}(k,l) \end{pmatrix} \qquad (8.1a)$$

$$\begin{pmatrix} \hat{\mathbf{d}}_1(k,l+1) \\ \hat{\mathbf{d}}_2(k,l+1) \\ \hat{\mathbf{d}}_3(k,l+1) \\ \hat{\mathbf{r}}(k,l+1) \end{pmatrix} = (\Delta t) \begin{pmatrix} (\Delta t)^{-1} & \omega_3 & -\omega_2 & -\lambda^2\, \gamma_1 \\ -\omega_3 & (\Delta t)^{-1} & -\omega_1 & -\lambda^2\, \gamma_2 \\ -\omega_2 & \omega_1 & (\Delta t)^{-1} & -\lambda^2\, \gamma_3 \\ \lambda^2\, \gamma_1 & \lambda^2\, \gamma_2 & \lambda^2\, \gamma_3 & (\Delta t)^{-1} \end{pmatrix} \begin{pmatrix} \hat{\mathbf{d}}_1(k,l) \\ \hat{\mathbf{d}}_2(k,l) \\ \hat{\mathbf{d}}_3(k,l) \\ \hat{\mathbf{r}}(k,l) \end{pmatrix}. \qquad (8.1b)$$

Eq.(8.1a) can be obtained from Eqs.(2.5) and (2.6) when the $O(\Delta s^2)$ is omitted. Eq.(8.1b) can be obtained from Eqs.(2.11) and (2.12) when the $O(\Delta t^2)$ is omitted.

If we assume zero elastic energy $h(\Omega,\Gamma)$ and zero kinetic energy $h(\omega,\gamma)$ in the constitutive relations (3.1a-3.10d), then $P = M = p = m = (0,0,0)^T$. Consequently the last two dynamic vector equations (3.13c, 3.13d) in the SMK equations (3.13) for force balance and torque balance become zero identically.

Furthermore, if we let
$$\hat{\mathbf{d}}_3 = \hat{\mathbf{t}},\ \hat{\mathbf{d}}_2 = \hat{\mathbf{n}},\ \hat{\mathbf{d}}_1 = \hat{\mathbf{b}} \qquad (8.2a)$$
$$\kappa = \Omega_1/\Gamma_3\ ,\ \tau = -\Omega_3/\Gamma_3\ . \qquad (8.2b)$$
$$\Omega_2 = \Gamma_1 = \Gamma_2 = 0, \qquad (8.2c)$$
then (8.1a) and (8.1b) become

$$\begin{pmatrix} \hat{\mathbf{b}}(k+1,l) \\ \hat{\mathbf{n}}(k+1,l) \\ \hat{\mathbf{t}}(k+1,l) \\ \hat{\mathbf{r}}(k+1,l) \end{pmatrix} = (\Delta\tilde{s}) \begin{pmatrix} (\Delta\tilde{s})^{-1} & -\tau & 0 & 0 \\ \tau & (\Delta\tilde{s})^{-1} & -\kappa & 0 \\ 0 & \kappa & (\Delta\tilde{s})^{-1} & -\lambda^2 \\ 0 & 0 & \lambda^2 & (\Delta\tilde{s})^{-1} \end{pmatrix} \begin{pmatrix} \hat{\mathbf{b}}(k,l) \\ \hat{\mathbf{n}}(k,l) \\ \hat{\mathbf{t}}(k,l) \\ \hat{\mathbf{r}}(k,l) \end{pmatrix} \qquad (8.3a)$$

$$\begin{pmatrix} \hat{\mathbf{b}}(k,l+1) \\ \mathbf{n}(k,l+1) \\ \mathbf{t}(k,l+1) \\ \hat{\mathbf{r}}(k,l+1) \end{pmatrix} = (\Delta t) \begin{pmatrix} (\Delta t)^{-1} & \omega_3 & -\omega_2 & -\lambda^2 \gamma_1 \\ -\omega_3 & (\Delta t)^{-1} & -\omega_1 & -\lambda^2 \gamma_2 \\ -\omega_2 & \omega_1 & (\Delta t)^{-1} & -\lambda^2 \gamma_3 \\ \lambda^2 \gamma_1 & \lambda^2 \gamma_2 & \lambda^2 \gamma_3 & (\Delta t)^{-1} \end{pmatrix} \begin{pmatrix} \hat{\mathbf{b}}(k,l) \\ \mathbf{n}(k,l) \\ \mathbf{t}(k,l) \\ \hat{\mathbf{r}}(k,l) \end{pmatrix} \quad (8.3b)$$

where $\Delta \tilde{s} = \Gamma_3 \Delta s$.

With $\tilde{s} = k \Delta \tilde{s}$ and $t = l \Delta t$, Eqs. (8.3a) and (8.3b) become the relations satisfied by the Serret-Frenet frame $\{\hat{\mathbf{n}}(\tilde{s},t,\lambda), \hat{\mathbf{b}}(\tilde{s},t,\lambda), \hat{\mathbf{t}}(\tilde{s},t,\lambda), \hat{\mathbf{r}}(\tilde{s},t,\lambda)\}$ when a discrete space curve moves on a real 3-dimensional spherical surface $S^3$ with radius $\lambda^{-2}$. In Eqs. (8.3a) and (8.3b), $\hat{\mathbf{r}}(\tilde{s},t,\lambda) = \lambda^{-2} \mathbf{r}(\tilde{s},t,\lambda)$ is the unit radius vector, $\kappa = \kappa(\tilde{s},t,\lambda)$ is the curvature, and $\tau = \tau(\tilde{s},t,\lambda)$ is the geometric torsion of the discrete space curve.

Since $\Gamma_3$ is a function of $(k,l)$, then $\Delta \tilde{s}$, the step size in arclength $\tilde{s}$, is a function of $(k,l)$ as well.

When $\Gamma_3$ becomes a constant, $\Delta \tilde{s}$ also becomes a constant. Doliwa and Santini have investigated this situation in [15].

Thus a discrete space curve with constant step size in arclength is a special case of the discrete elastic rod with no shear deformation ($\Gamma_1 = \Gamma_2 = 0$), with constant extension ($\Gamma_3 = $ constant), with no bending deformation in $\hat{\mathbf{d}}_2 = \hat{\mathbf{n}}$ direction ($\Omega_2 = 0$), with zero elastic energy function $H(\Omega,\Gamma)$, and with zero kinetic energy function $h(\omega,\gamma)$.

This is the key link connecting the s-and-t-discrete SMK equations (3.13), *via* the moving discrete curve problem (8.3a,b), with most well-known integrable discrete systems in dimension (1+1) (Ablowitz-Ladik etc) [16].

There also exists another tetrad frame $(\hat{\mathbf{e}}_1(s,t,\lambda), \hat{\mathbf{e}}_2(s,t,\lambda), \hat{\mathbf{e}}_3(s,t,\lambda), \hat{\mathbf{e}}_4(s,t,\lambda))$ which satisfies the following relations:

$$\begin{pmatrix} \hat{\mathbf{e}}_1(k+1,l) \\ \hat{\mathbf{e}}_2(k+1,l) \\ \hat{\mathbf{e}}_3(k+1,l) \\ \hat{\mathbf{e}}_4(k+1,l) \end{pmatrix} = (\Delta s) \begin{pmatrix} (\Delta s)^{-1} & p_3 & -p_2 & -\lambda^2 m_1 \\ -p_3 & (\Delta s)^{-1} & -p_1 & -\lambda^2 m_2 \\ -p_2 & p_1 & (\Delta s)^{-1} & -\lambda^2 m_3 \\ \lambda^2 m_1 & \lambda^2 m_2 & \lambda^2 m_3 & (\Delta s)^{-1} \end{pmatrix} \begin{pmatrix} \hat{\mathbf{e}}_1(k,l) \\ \hat{\mathbf{e}}_2(k,l) \\ \hat{\mathbf{e}}_3(k,l) \\ \hat{\mathbf{e}}_4(k,l) \end{pmatrix} \quad (8.4a)$$

$$\begin{pmatrix} \hat{\mathbf{e}}_1(k,l+1) \\ \hat{\mathbf{e}}_2(k,l+1) \\ \hat{\mathbf{e}}_3(k,l+1) \\ \hat{\mathbf{e}}_4(k,l+1) \end{pmatrix} = (\Delta t) \begin{pmatrix} (\Delta t)^{-1} & P_3 & -P_2 & -\lambda^2 M_1 \\ -P_3 & (\Delta t)^{-1} & -P_1 & -\lambda^2 M_2 \\ -P_2 & P_1 & (\Delta t)^{-1} & -\lambda^2 M_3 \\ \lambda^2 M_1 & \lambda^2 M_2 & \lambda^2 M_3 & (\Delta t)^{-1} \end{pmatrix} \begin{pmatrix} \hat{\mathbf{e}}_1(k,l) \\ \hat{\mathbf{e}}_2(k,l) \\ \hat{\mathbf{e}}_3(k,l) \\ \hat{\mathbf{e}}_4(k,l) \end{pmatrix}. \quad (8.4b)$$

Since the elements in 4-by-4 matrices in (8.2a,b) are related to the elements of the 4-by-4 matrices in (8.4a,b) *via* the constitutive relations like (3.10a-3.10d), we may say that the frame $\{\hat{\mathbf{e}}_1, \hat{\mathbf{e}}_2, \hat{\mathbf{e}}_3, \hat{\mathbf{e}}_4\}$ is dual to the Serret-Frenet frame $\{\tilde{\mathbf{d}}_1, \tilde{\mathbf{d}}_2, \tilde{\mathbf{d}}_3, \hat{\mathbf{r}}\}$.

# V. RELATION TO SPIN DESCRIPTION OF SOLITON EQUATIONS

We may rewrite the first component equation in (8.1b) as

$$\hat{\mathbf{S}}(k+1,l) = \hat{\mathbf{S}}(k,l) + \Delta s \sum_{J=2}^{4} b_J \hat{\mathbf{d}}_J(k,l), \qquad \hat{\mathbf{S}}(k,l) \equiv \hat{\mathbf{d}}_1(k,l). \tag{9.1}$$

This is an s-and-t-discrete version of the basic equation in Myrzakulov's unit spin description of the integrable and nonintegrable Partial-Differential-Equations [17].

Eqs.(8.1a,b) tell us that it is better to consider the motion of not just unit vector $\hat{\mathbf{S}} \equiv \hat{\mathbf{d}}_1$, but rather the motion of all unit vectors $\hat{\mathbf{d}}_J$ ($J=1,2,3,4$) together in (1+1) dimension. We conjecture that any system of integrable or nonintegrable Partial Difference Equations in (1+1) dimension derived from Eq.(9.1) might also be derived from a system of Partial Difference Equations in (3.9) with a Lax pair of ($\tilde{U}$ and $\tilde{V}$ in 3.3) or ($\tilde{X}$ and $\tilde{Y}$ in B.2).

If we use the 8-by-8 Lax pair as shown in Appendix B and choose the normalization factor properly for $\Psi$ in (B.1), then we can rewrite (B.1) as

$$\hat{\mathbf{f}}_I(k+1,l) = \hat{\mathbf{f}}_I(k,l) + \Delta s\, X_{IJ}\, \hat{\mathbf{f}}_J(k,l), \qquad (I=1,2,...,8), \tag{9.2a}$$

$$\hat{\mathbf{f}}_I(k,l+1) = \hat{\mathbf{f}}_I(k,l) + \Delta t\, Y_{IJ}\, \hat{\mathbf{f}}_J(k,l), \qquad (I=1,2,...,8), \tag{9.2b}$$

where $(\hat{\mathbf{f}}_1, \hat{\mathbf{f}}_2, ..., \hat{\mathbf{f}}_8)^T = \Psi^T$, $\hat{\mathbf{f}}_I \cdot \hat{\mathbf{f}}_J = c_J \delta_{IJ}$ ($c_J$ is a complex contant), and $X$ and $Y$ are the 8-by-8 matrices defined in (B.2a) and (B.2b) respectively.

Matrices $X$ and $Y$ have the following symmetry properties:

$$X_{IJ} = X_{JI}, \qquad Y_{IJ} = Y_{JI} \qquad \text{(if } I, J-4 = 1, 2, 3, 4 \text{ or } I-4, J = 1,2,3,4\text{)}, \tag{9.3a}$$

$$X_{IJ} = -X_{JI}, \qquad Y_{IJ} = -Y_{JI} \qquad \text{(if } I,J = 1,2,3,4 \text{ or } I,J = 5,6,7,8\text{)}. \tag{9.3b}$$

Since matrix $X$ is not antisymmetric, Eq. (9.2a) is not a Serret-Frenet equation for an elastic rod moving on a 7D sphere ($S^7$) with radius $|\zeta|^{-2} = \lambda^{-2}$ imbedded in $R^8$.

Because the diagonal matrix elements of $Y$ are all zero, we may rewrite the first component equation in (9.2b) as

$$\hat{\mathbf{S}}(k+1,l) = \hat{\mathbf{S}}(k,l) + \Delta s \sum_{J=2}^{8} a_J(k,l)\, \hat{\mathbf{f}}_J(k,l), \qquad \hat{\mathbf{S}}(k,l) \equiv \hat{\mathbf{f}}_1(k,l) \tag{9.4}$$

where $a_J(k,l) \equiv Y_{1J}(k,l)$.

This is the s-and-t-discrete analog of Eq. (9.1) in 8D[18]. Similarly it is better to consider the motion of not just unit vector $\hat{\mathbf{S}} \equiv \hat{\mathbf{f}}_1$, but rather the motion of all unit vectors $\hat{\mathbf{f}}_J$ ($J=1,2,...,7,8$) together in (1+1) dimension.

# VI. CONCLUSIONS

(1) We have found a Lax pair with a corresponding spectral parameter for a system of 12 scalar Partial-Difference-Equations and 12 scalar constitutive relations governing 24 dependent variables in (1+1) dimension.
(2) When the spectral parameter goes to zero, this system of Partial-Difference-Equations reduces to the s-and-t-discrete SMK equations that describe the discrete dynamics of DNA modeled as a shearable and extensible discrete elastic rod.
(3) When three dependent variables are set to be constants, the s-and-t-discrete SMK equations reduce to a system of 9 scalar Partial-Difference-Equations with 9 scalar constitutive relations in 18 dependent variables. This system describes the discrete dynamics of DNA modeled as an unshearable and inextensible discrete elastic rod (discrete Kirchhoff elastic rod).
(4) When the s-and-t-discrete SMK equations are assumed to be independent of $t = l \Delta t$ or $s = k \Delta s$, they reduce to a set of 6 Difference-Equations and 3 constitutive relations for 9 dependent variables describing the discrete motion of a heavy top, or the discrete motion of a rigid body in an ideal fluid.

# APPENDIX A

Figure 1[19], The following Figure is copied from
(http://biop.ox.ac.uk/www/lab_journal_1998/Dickerson/FIG1.gif)

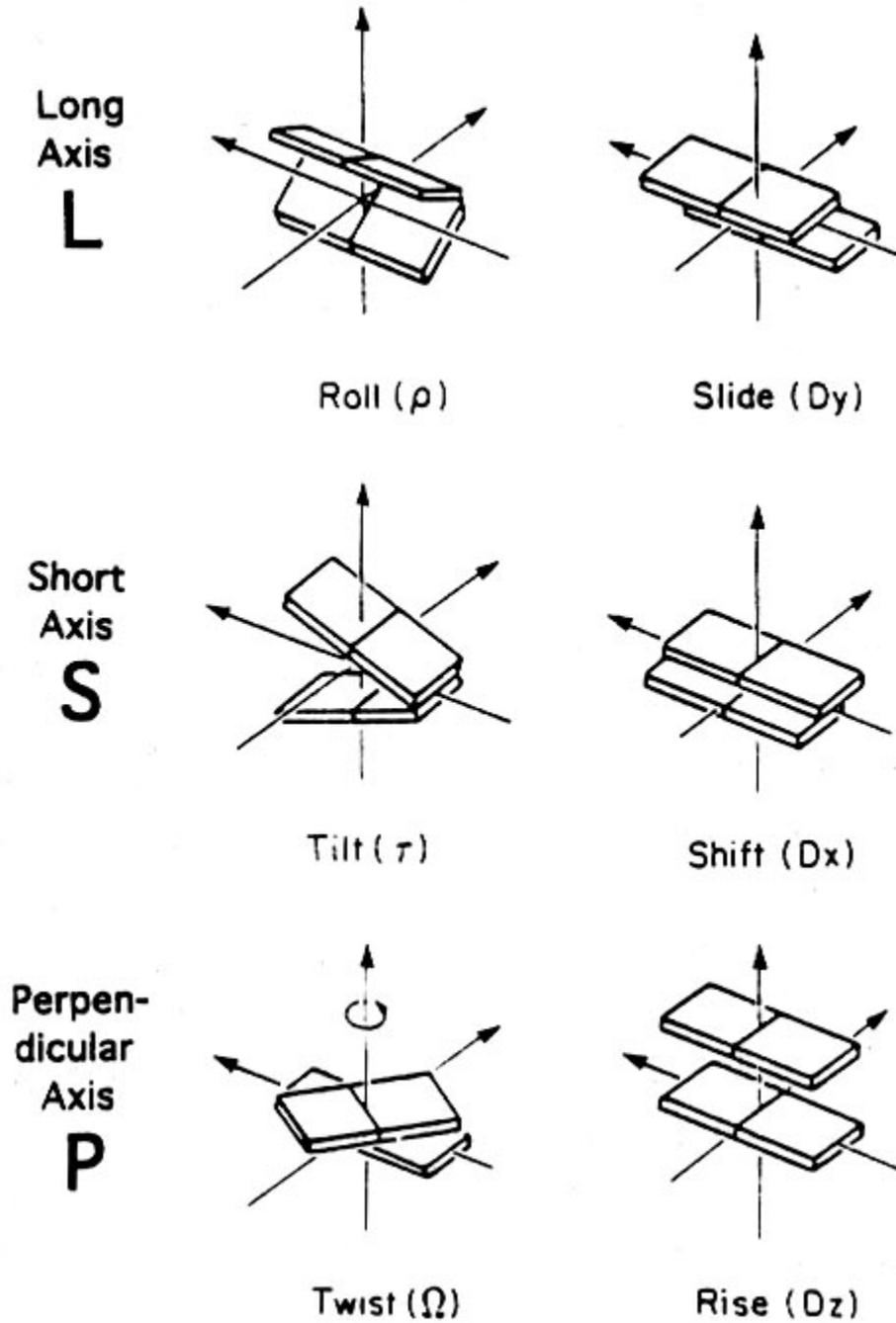

Table IA. The basepair-averaged values of the intrinsic Roll, Tilt, Twist, Slide, Shift, and Rise of the naked B-DNA.

| Twist($\Omega_3^{(intrinsic)}\Delta s$) | Tilt($\Omega_2^{(intrinsic)}\Delta s$) | Roll($\Omega_1^{(intrinsic)}\Delta s$) |
|---|---|---|
| 35.58668 deg | -0.70584 deg | 2.559459 deg |

| Shift($\Gamma_2^{(intrinsic)}\Delta s$) | Slide($\Gamma_1^{(intrinsic)}\Delta s$) | Rise($\Gamma_3^{(intrinsic)}\Delta s$) |
|---|---|---|
| 0.00171 nm | -0.001474 nm | 0.3335395 nm |

Table IB[20]. Average values and dispersion (listed below average) of base pair step parameters in naked B-DNA (http://rutchem.rutgers.edu/~olson/ave_dpn.html)

```
                         Part B: B-DNA

   Step    N    Twist(°)  Tilt(°)  Roll(°)  Shift(Å)  Slide(Å)  Rise(Å)

   AA     81     35.5     -0.8      0.1     -0.00     -0.14      3.28
                  3.6      3.1      3.8      0.39      0.30      0.17
   AG     26     30.6     -2.4      3.6      0.27      0.25      3.24
                  4.7      2.9      2.4      0.37      0.48      0.15
   GA     45     39.6     -1.1      0.5      0.02     -0.06      3.39
                  3.0      2.9      3.7      0.34      0.32      0.17
   GG      5     35.3     -2.0      5.2     -0.17      0.64      3.41
                  4.9      2.6      2.8      0.59      0.08      0.13

   AC     14     33.1     -1.3     -0.3     -0.30     -0.20      3.28
                  4.6      3.6      4.6      0.53      0.46      0.17
   AT     94     31.6     -0.0     -0.9      0.00     -0.49      3.34
                  3.4      2.5      3.5      0.39      0.20      0.18
   GC     96     38.4      0.0     -6.0      0.00      0.38      3.54
                  3.0      3.8      4.3      0.88      0.27      0.15

   CA     37     37.7      0.2      1.7     -0.01      1.47      3.26
                  9.3      2.9      6.2      0.36      0.96      0.18
   CG     56     31.3      0.0      6.2     -0.00      0.82      3.23
                  4.7      3.5      3.7      0.47      0.27      0.23
   TA     34     43.2     -0.0     -0.1      0.00      0.87      3.43
                  5.5      3.2      4.2      0.39      1.00      0.14
```

Table IIA. The matrices $A_{ab}$, $B_{ab}$, and, $C_{ab}$ are the sub-matrices of the so called covariance matrix. For naked B-DNA, their basepair-averaged values are in the square brackets[].

| | Twist | Tilt | Roll | Shift | Slide | Rise |
|---|---|---|---|---|---|---|

| Twist | $A_{33}$[24.876] | $A_{32}$[-1.072] | $A_{31}$[-10.678] | $B_{32}$[-0.257] | $B_{31}$[1.238] | $B_{33}$[0.18] |
|---|---|---|---|---|---|---|
| Tilt | $A_{23}$[-1.072] | $A_{22}$[9.714] | $A_{21}$[0.125] | $B_{22}$[0.3] | $B_{21}$[-0.165] | $B_{23}$[0.083] |
| Roll | $A_{13}$[-10.678] | $A_{12}$[0.125] | $A_{11}$[16.27] | $B_{12}$[0.254] | $B_{11}$[-0.775] | $B_{13}$[-0.143] |
| Shift | $B_{32}$[-0.257] | $B_{22}$[0.3] | $B_{12}$[0.254] | $C_{22}$[0.247] | $C_{21}$[-0.002] | $C_{23}$[-0.01] |
| Slide | $B_{31}$[1.238] | $B_{21}$[-0.165] | $B_{11}$[-0.775] | $C_{12}$[-0.002] | $C_{11}$[0.275] | $C_{13}$[-0.005] |
| Rise | $B_{33}$[0.18] | $B_{23}$[0.083] | $B_{13}$[-0.143] | $C_{32}$[-0.01] | $C_{31}$[-0.005] | $C_{33}$[0.028] |

Table IIB[21]. Covariance Matrix for Dimer Steps in Naked B-DNA
(http://rutchem.rutgers.edu/~olson/cov_matrix.html)

```
                            Part B: Naked B-DNA

                                  CG
             Twist      Tilt      Roll     Shift     Slide      Rise
   Twist     21.98     -0.00     -3.37      0.00     -0.36      0.58
    Tilt     -0.00     12.10      0.00     -0.18      0.00     -0.00
    Roll     -3.37      0.00     13.80      0.00     -0.21     -0.18
   Shift      0.00     -0.18      0.00      0.23     -0.00      0.00
   Slide     -0.36      0.00     -0.21     -0.00      0.07     -0.01
    Rise      0.58     -0.00     -0.18      0.00     -0.01      0.05

                                  CA
             Twist      Tilt      Roll     Shift     Slide      Rise
   Twist     86.10     -6.66    -49.41      1.36      7.35      0.52
    Tilt     -6.66      8.59      2.39     -0.16     -0.37     -0.20
    Roll    -49.41      2.39     37.91     -0.64     -5.09     -0.20
   Shift      1.36     -0.16     -0.64      0.13      0.15      0.01
   Slide      7.35     -0.37     -5.09      0.15      0.91      0.02
    Rise      0.52     -0.20     -0.20      0.01      0.02      0.03

                                  TA
             Twist      Tilt      Roll     Shift     Slide      Rise
   Twist     30.63     -0.00    -16.66      0.00      4.58     -0.01
    Tilt     -0.00     10.47      0.00      0.29     -0.00      0.00
    Roll    -16.66      0.00     17.41     -0.00     -1.82     -0.04
   Shift      0.00      0.29     -0.00      0.15      0.00      0.00
   Slide      4.58     -0.00     -1.82      0.00      1.01     -0.05
    Rise     -0.01      0.00     -0.04      0.00     -0.05      0.02

                                  AG
             Twist      Tilt      Roll     Shift     Slide      Rise
   Twist     22.13     -0.42     -2.56     -0.23     -1.60      0.03
    Tilt     -0.42      8.24     -2.80     -0.52     -0.40      0.23
    Roll     -2.56     -2.80      5.59      0.51      0.55     -0.19
   Shift     -0.23     -0.52      0.51      0.13      0.09     -0.03
   Slide     -1.60     -0.40      0.55      0.09      0.23     -0.03
    Rise      0.03      0.23     -0.19     -0.03     -0.03      0.02

                                  GG
```

|       | Twist | Tilt  | Roll  | Shift | Slide | Rise  |
|-------|-------|-------|-------|-------|-------|-------|
| Twist | 24.27 | -2.65 | -9.37 | -2.34 |  0.03 |  0.33 |
| Tilt  | -2.65 |  6.58 |  0.73 | -0.21 |  0.01 |  0.25 |
| Roll  | -9.37 |  0.73 |  8.00 |  1.36 | -0.17 | -0.16 |
| Shift | -2.34 | -0.21 |  1.36 |  0.35 | -0.02 | -0.06 |
| Slide |  0.03 |  0.01 | -0.17 | -0.02 |  0.01 |  0.00 |
| Rise  |  0.33 |  0.25 | -0.16 | -0.06 |  0.00 |  0.02 |

AA

|       | Twist | Tilt  | Roll  | Shift | Slide | Rise  |
|-------|-------|-------|-------|-------|-------|-------|
| Twist | 13.20 |  0.18 | -4.67 | -0.22 |  0.08 |  0.13 |
| Tilt  |  0.18 |  9.32 | -1.46 | -0.15 | -0.28 |  0.22 |
| Roll  | -4.67 | -1.46 | 14.75 |  0.33 |  0.31 | -0.01 |
| Shift | -0.22 | -0.15 |  0.33 |  0.15 |  0.00 |  0.00 |
| Slide |  0.08 | -0.28 |  0.31 |  0.00 |  0.09 | -0.00 |
| Rise  |  0.13 |  0.22 | -0.01 |  0.00 | -0.00 |  0.03 |

GA

|       | Twist | Tilt  | Roll  | Shift | Slide | Rise  |
|-------|-------|-------|-------|-------|-------|-------|
| Twist |  9.16 | -0.19 | -4.38 |  0.09 |  0.21 |  0.08 |
| Tilt  | -0.19 |  8.14 | -3.44 |  0.37 | -0.11 |  0.29 |
| Roll  | -4.38 | -3.44 | 13.44 | -0.22 |  0.02 | -0.21 |
| Shift |  0.09 |  0.37 | -0.22 |  0.12 | -0.05 | -0.01 |
| Slide |  0.21 | -0.11 |  0.02 | -0.05 |  0.10 |  0.01 |
| Rise  |  0.08 |  0.29 | -0.21 | -0.01 |  0.01 |  0.03 |

AT

|       | Twist | Tilt  | Roll  | Shift | Slide | Rise  |
|-------|-------|-------|-------|-------|-------|-------|
| Twist | 11.39 |  0.00 | -6.09 | -0.00 | -0.09 |  0.15 |
| Tilt  |  0.00 |  6.35 | -0.00 |  0.16 |  0.00 |  0.00 |
| Roll  | -6.09 | -0.00 | 12.00 |  0.00 |  0.06 | -0.15 |
| Shift | -0.00 |  0.16 |  0.00 |  0.15 | -0.00 | -0.00 |
| Slide | -0.09 |  0.00 |  0.06 | -0.00 |  0.04 | -0.00 |
| Rise  |  0.15 |  0.00 | -0.15 | -0.00 | -0.00 |  0.03 |

AC

|       | Twist | Tilt  | Roll  | Shift | Slide | Rise  |
|-------|-------|-------|-------|-------|-------|-------|
| Twist | 21.03 | -0.98 | -6.01 | -1.23 |  1.73 | -0.03 |
| Tilt  | -0.98 | 12.83 |  5.83 |  1.41 | -0.50 |  0.04 |
| Roll  | -6.01 |  5.83 | 21.35 |  1.20 | -0.92 | -0.16 |
| Shift | -1.23 |  1.41 |  1.20 |  0.28 | -0.19 | -0.01 |
| Slide |  1.73 | -0.50 | -0.92 | -0.19 |  0.22 |  0.02 |
| Rise  | -0.03 |  0.04 | -0.16 | -0.01 |  0.02 |  0.03 |

GC

|       | Twist | Tilt  | Roll  | Shift | Slide | Rise  |
|-------|-------|-------|-------|-------|-------|-------|
| Twist |  8.87 |  0.00 | -4.26 |  0.00 |  0.45 |  0.02 |
| Tilt  |  0.00 | 14.52 | -0.00 |  1.99 |  0.00 |  0.00 |
| Roll  | -4.26 | -0.00 | 18.45 | -0.00 | -0.48 | -0.13 |
| Shift |  0.00 |  1.99 | -0.00 |  0.78 |  0.00 |  0.00 |
| Slide |  0.45 |  0.00 | -0.48 |  0.00 |  0.07 | -0.01 |
| Rise  |  0.02 |  0.00 | -0.13 |  0.00 | -0.01 |  0.02 |

# APPENDIX B
# A different derivation of Eqs.(3.9a-d)

If we assume that 24 dependent variables $\Omega_a$, $\Gamma_a$, $\omega_a$, $\gamma_a$, $M_a$, $P_a$, $m_a$, $p_a$ ($a=1,2,3$) are complex and the parameter $\zeta$ is also complex, then we may consider the following linear system:

$$\Psi^1 = \tilde{X}\, \Psi \tag{B.1a}$$
$$\Psi_1 = \tilde{Y}\, \Psi \tag{B.1b}$$

where $\tilde{X}$ and $\tilde{Y}$ in the Lax pair $(\tilde{X},\tilde{Y})$ are defined as

$$\tilde{X} = \mathbf{1}_8 + \Delta s\, X \tag{B.2a}$$
$$\tilde{Y} = \mathbf{1}_8 + \Delta t\, Y \tag{B.2b}$$
$$X = \begin{pmatrix} A & -B \\ B & A \end{pmatrix} \tag{B.2c}$$
$$Y = \begin{pmatrix} C & -D \\ D & C \end{pmatrix} \tag{B.2d}$$

and where $\mathbf{1}_8$ is the 8-by-8 unit matrix and $A, B, C, D$ are given by

$$A = -\Omega_a\, J_a + \zeta^2\, \Gamma_a\, K_a \tag{B.2e}$$
$$B = \zeta\left(-p_a\, J_a + \zeta^2\, m_a\, K_a\right) \tag{B.2f}$$
$$C = -\omega_a\, J_a + \zeta^2\, \gamma_a\, K_a \tag{B.2g}$$
$$D = \zeta\left(-P_a\, J_a + \zeta^2\, M_a\, K_a\right). \tag{B.2h}$$

The integrability condition for system (B.1) leads to:
$$\tilde{X}_1\, \tilde{Y} = \tilde{Y}^1\, \tilde{X}. \tag{B.3}$$

Taking each 4-by-4 block in (B.3) and left- or right-multiplying it by $J_a$ and $K_a$ ($a=1,2,3$) respectively, using (3.1) to simplify the result, we obtain a set of Partial-Difference-Equations:

$$(\Delta t)^{-1}\left(\Gamma_{c,1} - \Gamma_c\right) - (\Delta s)^{-1}\left(\gamma_c^1 - \gamma_c\right)$$
$$-\varepsilon_{abc}\left(\Gamma_a\, \omega_b^1 + \Omega_{a,1}\, \gamma_b\right) + (\zeta)^2\, \varepsilon_{abc}\left(M_{a,1}\, p_{b,1} - m_a\, P_b^1\right) = 0 \tag{B.4a}$$

$$(\Delta t)^{-1}\left(\Omega_{a,1} - \Omega_a\right) - (\Delta s)^{-1}\left(\omega_a^1 - \omega_a\right)$$
$$-\left(\Omega_{a+1}\, \omega_{a+2}^1 - \omega_{a+1}\, \Omega_{a+2,1}\right) + (\zeta)^2\left(P_{a+1}\, p_{a+2,1} - p_{a+1}\, P_{a+2}^1\right) \tag{B.4b}$$
$$+(\zeta)^4\left(\Gamma_{a+1,1}\, \gamma_{a+2} - \gamma_{a+1}^1\, \Gamma_{a+2}\right) - (\zeta)^6\left(M_{a+1}^1\, m_{a+2} - m_{a+1,1}\, M_{a+2}\right) = 0$$

$$(\Delta t)^{-1}(p_{a,1} - p_a) - (\Delta s)^{-1}(P_a^1 - P_a)$$
$$- (p_{a+1}\, \omega_{a+2}^1 - \omega_{a+1}\, p_{a+2,1}) + (P_{a+1}\, \Omega_{a+2,1} - \Omega_{a+1}\, P_{a+2}^1) \quad \text{(B.4c)}$$
$$+ (\zeta)^4 (\Gamma_{a+1}\, M_{a+2}^1 - M_{a+1}\, \Gamma_{a+2,1}) + (\zeta)^4 (\gamma_{a+1}^1\, m_{a+2} - m_{a+1,1}\, \gamma_{a+2}) = 0$$

$$(\Delta t)^{-1}(m_{c,1} - m_c) - (\Delta s)^{-1}(M_c^1 - M_c)$$
$$- \varepsilon_{abc}(p_{a,1}\, \gamma_b + \Gamma_a\, P_b^1 + \Omega_{a,1}\, M_b + m_a\, \omega_b^1) = 0 \quad \text{(B.4d)}$$

In (B.4) above, index $a$ runs 1,2,3. Index $a+1$ and index $a+2$ are assumed to be taking the values of modula 3.

Assuming that $\zeta$ is pure imaginary, we find that system (B.4) reduces to system (3.9).

# APPENDIX C. SOLUTION TO THE SMK EQUATIONS

The s-and-t-discrete SMK equations (3.13) read:

$$(\Delta t)^{-1}(\Gamma_{c,1} - \Gamma_c) + \varepsilon_{abc}\omega_a^1 \Gamma_b$$
$$= (\Delta s)^{-1}(\gamma_c^1 - \gamma_c) + \varepsilon_{abc}\Omega_{a,1}\gamma_b \tag{C.1a}$$

$$(\Delta t)^{-1}(\Omega_{a,1} - \Omega_a) - \omega_{a+1}\Omega_{a+2,1}^1$$
$$= (\Delta s)^{-1}(\omega_a^1 - \omega_a) - \Omega_{a+1}\omega_{a+2}^1 \tag{C.1b}$$

$$(\Delta t)^{-1}(p_{a,1} - p_a) + (\omega_{a+1}p_{a+2,1} - p_{a+1}\omega_{a+2}^1)$$
$$= (\Delta s)^{-1}(P_a^1 - P_a) + (\Omega_{a+1}P_{a+2}^1 - P_{a+1}\Omega_{a+2,1}) \tag{C.1c}$$

$$(\Delta t)^{-1}(m_{c,1} - m_c) + \varepsilon_{abc}(\omega_a^1 m_b + \gamma_a p_{b,1})$$
$$= (\Delta s)^{-1}(M_c^1 - M_c) + \varepsilon_{abc}(\Omega_{a,1}M_b + \Gamma_a P_b^1) \tag{C.1d}$$

In (C.1) above, index $a$ runs 1,2,3. Index $a+1$ and index $a+2$ are assumed to take the values of modula 3.

The constitutive relations are

$$P_a = \frac{\partial}{\partial \Gamma_a} H(\Omega, \Gamma) \tag{C.1e}$$

$$M_a = \frac{\partial}{\partial \Omega_a} H(\Omega, \Gamma) \tag{C.1f}$$

$$p_a = \frac{\partial}{\partial \gamma_a} h(\omega, \gamma) \tag{C.1g}$$

$$m_a = \frac{\partial}{\partial \omega_a} h(\omega, \gamma) \tag{C.1h}$$

In the following we will express the $k$ and $l$ dependence explicitly.

$$(\Delta t)^{-1}(\Gamma_c(k,l+1) - \Gamma_c(k,l)) + \varepsilon_{abc}\omega_a(k+1,l)\Gamma_b(k,l)$$
$$= (\Delta s)^{-1}(\gamma_c(k+1,l) - \gamma_c(k,l)) + \varepsilon_{abc}\Omega_a(k,l+1)\gamma_b(k,l) \tag{C.2}$$

$$(\Delta t)^{-1}\left(\Omega_a(k,l+1)-\Omega_a(k,l)\right)-\omega_{a+1}(k,l)\Omega_{a+2,1}(k,l+1)$$
$$=(\Delta s)^{-1}\left(\omega_a(k+1,l)-\omega_a(k,l)\right)-\Omega_{a+1}(k,l)\,\omega_{a+2}(k+1,l) \quad \text{(C.3)}$$

$$(\Delta t)^{-1}\left(p_a(k,l+1)-p_a(k,l)\right)+\left(\omega_{a+1}(k,l)p_{a+2,1}(k,l+1)-p_{a+1}(k,l)\omega_{a+2}(k+1,l)\right)$$
$$=(\Delta s)^{-1}\left(P_a(k+1,l)-P_a(k,l)\right)+\left(\Omega_{a+1}(k,l)P^1_{a+2}(k+1,l)-P_{a+1}(k,l)\Omega_{a+2}(k,l+1)\right) \quad \text{(C.4)}$$

$$(\Delta t)^{-1}\left(m_c(k,l+1)-m_c(k,l)\right)+\varepsilon_{abc}\left(\omega_a(k+1,l)m_b(k,l)+\gamma_a(k,l)p_b(k,l+1)\right)$$
$$=(\Delta s)^{-1}\left(M^1_c(k+1,l)-M_c(k,l)\right)+\varepsilon_{abc}\left(\Omega_a(k,l+1)M_b(k,l)+\Gamma_a(k,l)P_b(k+1,l)\right) \quad \text{(C.5)}$$

We are going to assume that periodic boundary conditions apply. Therefore if $X(k,l)$ is known for all $k$, so is $X(k+1,l)$. The goal is then to express unknown quantities $X(k,l+1)$ in terms of the known quantities $X(k,l)$ and $X(k+1,l)$. When the initial conditions $X(k,l=0)$ and $X(k+1,l=0)$ are specified for all $k$, we can use obtain solutions for $X(k,l=1)$ and $X(k+1,l=1)$ and so on so forth.

Here are the detailed steps:
(0) Pick a value for $\Delta t$ such that $(\Delta t)^{-3} > |\omega_1(k,l)\omega_2(k,l)\omega_3(k,l)|$.

(1) Solve $\Omega_a(k,l+1)$ from Eq.(C.3) via:

$$f_a(k,l) = \Omega_a(k,l)$$
$$-(\Delta t)\Omega_{a+1}(k,l)\,\omega_{a+2}(k+1,l) \quad \text{(C.6a)}$$
$$+(\Delta t)(\Delta s)^{-1}\left(\omega_a(k+1,l)-\omega_a(k,l)\right)$$

$$\begin{pmatrix}\Omega_1(k,l+1)\\ \Omega_{2,1}(k,l+1)\\ \Omega_{3,1}(k,l+1)\end{pmatrix} = \begin{pmatrix} 1 & 0 & -\omega_2(k,l)\Delta t \\ -\omega_3(k,l)\Delta t & 1 & 0 \\ 0 & -\omega_1(k,l)\Delta t & 1 \end{pmatrix}^{-1} \begin{pmatrix} f_1(k,l)\\ f_2(k,l)\\ f_3(k,l)\end{pmatrix} \quad \text{(C.6b)}$$

(2) Solve $p_a(k,l+1)$ from Eq.(C.4) using solution $\Omega_a(k,l+1)$ via:

$$g_a(k,l) = p_a(k,l)$$
$$+ (\Delta t) p_{a+1}(k,l) \omega_{a+2}(k+1,l)$$
$$+ (\Delta t)(\Delta s)^{-1} \left( P_a(k+1,l) - P_a(k,l) \right) \quad \text{(C.7a)}$$
$$+ (\Delta t) \left( \Omega_{a+1}(k,l) P^1_{a+2}(k+1,l) - P_{a+1}(k,l) \Omega_{a+2}(k,l+1) \right)$$

$$\begin{pmatrix} p_1(k,l+1) \\ p_2(k,l+1) \\ p_3(k,l+1) \end{pmatrix} = \begin{pmatrix} 1 & 0 & \omega_2(k,l)\Delta t \\ \omega_3(k,l)\Delta t & 1 & 0 \\ 0 & \omega_1(k,l)\Delta t & 1 \end{pmatrix}^{-1} \begin{pmatrix} g_1(k,l) \\ g_2(k,l) \\ g_3(k,l) \end{pmatrix} \quad \text{(C.7b)}$$

(3) Solve $\Gamma_c(k,l+1)$ from Eq.(C.2) by using solution $\Omega_a(k,l+1)$ via:

$$\Gamma_c(k,l+1) = \Gamma_c(k,l)$$
$$+ (\Delta t)(\Delta s)^{-1} \left( \gamma_c(k+1,l) - \gamma_c(k,l) \right) \quad \text{(C.8)}$$
$$+ (\Delta t) \varepsilon_{abc} \left( \Omega_a(k,l+1) \gamma_b(k,l) - \omega_a(k+1,l) \Gamma_b(k,l) \right)$$

(4) Solve $m_c(k,l+1)$ from Eq.(C.5) by using solutions $\Omega_a(k,l+1)$ and $p_a(k,l+1)$ via:

$$m_c(k,l+1) = m_c(k,l)$$
$$+ (\Delta t)(\Delta s)^{-1} \left( M_c(k+1,l) - M_c(k,l) \right)$$
$$+ (\Delta t) \varepsilon_{abc} \left( \Omega_a(k,l+1) M_b(k,l) + \Gamma_a(k,l) P_b(k,l) \right) \quad \text{(C.9)}$$
$$- (\Delta t) \varepsilon_{abc} \left( \omega_a(k+1,l) m_b(k,l) + \gamma_a(k,l) p_b(k,l+1) \right)$$

(5) Go to step (1).

End of Appendix C.